\documentclass[apj,twocolumn]{emulateapj}
\eqsecnum

\renewcommand\email\texttt

\def\spose#1{\hbox to 0pt{#1\hss}}
\def\lta{\mathrel{\spose{\lower 3pt\hbox{$\sim$}}
    \raise 2.0pt\hbox{$<$}}}
\def\gta{\mathrel{\spose{\lower 3pt\hbox{$\sim$}}
    \raise 2.0pt\hbox{$>$}}}

\def\zl{z_{\rm L}}
\def\zs{z_{\rm S}}

\def\name{Cosmic Horseshoe}

\begin{document} 

\slugcomment{\sc submitted to \it the Astrophysical Journal}
\shorttitle{\sc The Cosmic Horseshoe} 
\shortauthors{Belokurov et al.}

\title{The Cosmic Horseshoe: Discovery of an Einstein Ring around a
  Giant Luminous Red Galaxy}

\author{V. Belokurov\altaffilmark{1},
N.~W.~Evans\altaffilmark{1}, 
A.~Moiseev\altaffilmark{2},  
L.~J.~King\altaffilmark{1},
P.~C.~Hewett\altaffilmark{1},
M.~Pettini\altaffilmark{1},
L.~Wyrzykowski\altaffilmark{1,4},  
R.~G.~McMahon\altaffilmark{1}, 
M.~C.~Smith\altaffilmark{1}, 
G.~Gilmore\altaffilmark{1}, 
S.~F.~Sanchez\altaffilmark{3}, 
A.~Udalski\altaffilmark{4}, 
S.~Koposov\altaffilmark{5},
D.~B.~Zucker\altaffilmark{1},
C.~J.~Walcher\altaffilmark{6}
}

\altaffiltext{1}{Institute of Astronomy, University of Cambridge,
Madingley Road, Cambridge CB3 0HA, UK;\email{vasily,nwe@ast.cam.ac.uk}}
\altaffiltext{2}{Special Astrophysical Observatory, Nizhniy Arkhyz,
Karachaevo-Cherkessiya, Russia;\email{moisav@sao.ru}}
\altaffiltext{3}{CAHA de Calar Alto (CSIC-MPIA), E4004 Almería, Spain}
\altaffiltext{4}{Warsaw University Observatory, Al. Ujazdowskie 4,
  00-479, Poland}
\altaffiltext{5}{MPIA, K\"{o}nigstuhl
17, 69117 Heidelberg, Germany}
\altaffiltext{6}{CNRS-Universit\'{e} de Provence, BP8, 
13376 Marseille Cedex 12, France}

\begin{abstract}
We report the discovery of an almost complete ($\sim 300^\circ$)
Einstein ring of diameter $10^{\prime\prime}$ in Sloan Digital Sky
Survey (SDSS) Data Release 5 (DR5).  Spectroscopic data from the 6m
telescope of the Special Astrophysical Observatory reveals that the
deflecting galaxy has a line-of-sight velocity dispersion in excess of
$400$ kms$^{-1}$ and a redshift of 0.444, whilst the source is a
star-forming galaxy with a redshift of 2.379. From its color,
luminosity and velocity dispersion, we argue that this is the most
massive galaxy lens hitherto discovered.
\end{abstract}


\section{Introduction}

There have been many optical giant arcs discovered, caused by the
lensing effects of massive galaxy clusters and their central
galaxies. But, few optical rings have ever been found, despite
theoretical predictions that they should be
abundant~\citep{Mi92}. \citet{Wa96} found 0047-2808, which is a high
redshift star-forming galaxy lensed by a massive early-type galaxy
into a partial ($\sim 170^\circ$) Einstein ring of $2\farcs70$
diameter, while \citet{Ki98} discovered a complete Einstein ring in
the near-infrared around B1938+666. \citet{Ca05} found a nearly
complete ($\sim 260^\circ$) Einstein ring of diameter $2\farcs96$
produced by the lensing of a starburst galaxy by a massive and
isolated elliptical galaxy. The Sloan Lens ACS
Survey~\citep[SLACS,][]{Bo06} has also identified a number of partial
optical rings, based on the identification of anomalous emission lines
in Sloan Digital Sky Survey (SDSS) spectra together with confirmatory
follow-up from the Advanced Camera for Surveys. Here, we report the
discovery of the \name\ -- an almost complete ($\sim 300^\circ$),
giant Einstein ring, with a diameter of $\sim 10^{\prime\prime}$ in
SDSS data. Throughout the paper, we use the standard cosmology
$\Omega_{\rm m}=0.3$, $\Omega_{\Lambda}=0.7$, $H_{0}=70$ km s$^{-1}$
Mpc$^{-1}$.

\begin{figure}
\begin{center}
\includegraphics[height=6cm]{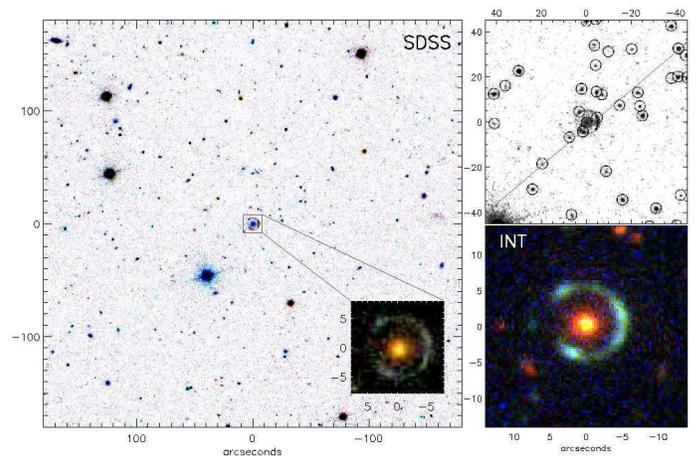}
\caption{\label{fig:image} Left: SDSS view of the sky composed from
  $g,r,i$ images around the \name. Most of the objects in the field
  are faint galaxies. The inset shows $16'' \times 16''$ cut-out
  centered on the lens. Note the bluish color of the ring.  Top right:
  SDSS $g,r,i$ composite with objects detected by the SDSS pipeline
  marked with circles. We also show the slit position for SAO
  follow-up. Bottom right: INT $u,g,i$ composite from follow-up data.}
\end{center}
\end{figure}

\begin{figure}
\begin{center}
\includegraphics[height=6cm]{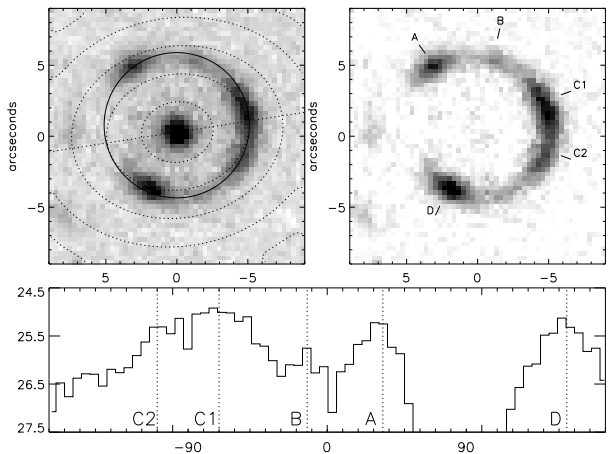} 
\caption{\label{fig:int} Left: $g$ band INT images of a $18'' \times
  18''$ field of view centered on the \name.  Dotted lines mark the
  major axis of the LRG and contours show isophotes at 1,2,3,4,5
  $R_{\rm eff}$ along the major axis. The best fit circle through the
  ring is shown as a solid line. Right: Decomposition of the light
  into the ring after subtraction of the luminosity model for the
  LRG. Also shown is the profile along the ring in the inset. The
  locations of the four maxima are marked.}
\end{center}
\end{figure}

\section{Discovery and Follow-Up}

Previous search strategies with SDSS data can be divided into three
kinds.  The first discovery was made by \citet{In03a}, who searched
around spectroscopically identified quasars looking for stellar-like
objects with a similar color to provide candidates for follow-up, and
found the spectacular $14\farcs62$ separation lens SDSS J1004+4112.
The remaining two methods target smaller separation lenses, in which
the images are unresolved by SDSS.  \citet{In03b} and \citet{Jo03}
searched through spectroscopically identified quasars, looking for
evidence of extended sources corresponding to unresolved, multiple
images.  The most widely-used strategy is to search through the
spectroscopic database looking for emission lines of high redshift
objects within the spectrum of lower redshift early-type galaxies
~\citep{Wi05, Bo06}. Here, we introduce a new method, inspired by the
recent, serendipitous discovery of the 8 O'clock Arc, which is a Lyman
Break galaxy lensed into three images merging into an extended
arc~\citep{Al07}. The SDSS pipeline resolved the arc into three
objects. This suggests searching for multiple, blue, faint companions
around luminous red galaxies (LRGs) in the SDSS object catalogue. The
search is fast, so it is easy to experiment with different magnitude
and color cuts, as well as search radii. For example, selecting lenses
in DR5 to be brighter than $r = 19.5$ and $g\!-\!r > 0.6$, together
with sources within $6^{\prime\prime}$ that are fainter than $r =
19.5$ and bluer than $g\!-\!r = 0.5$ yields 3 very strong candidates.
One of the three candidates is the 8 O'clock arc -- another is the
subject of this {\it Letter}, the \name.

\begin{figure*}
\begin{center}
\includegraphics[height=4cm]{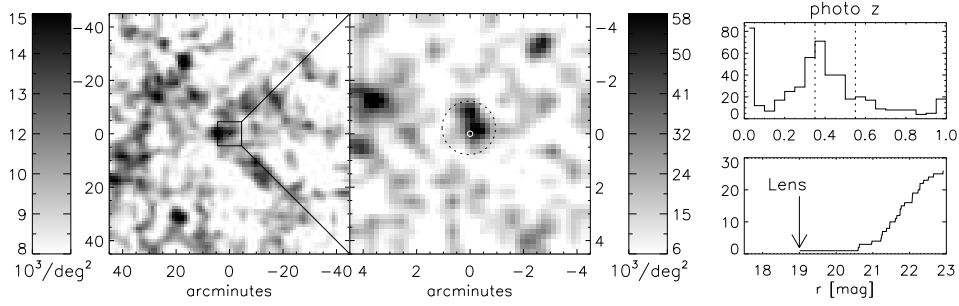} 
\caption{\label{fig:nbhd} Density of galaxies in the vicinity of the
  object with SDSS photometric redshifts in the range $0.35 < z <
  0.55$. Left: Large scale structure. Middle: Zoom-in on the lens
  marked by white ring, shown to scale. The lens belongs to the group
  of $\sim 26$ galaxies, marked by dashed circle of $1'$
  radius. Right: Redshift distribution (upper panel) for all galaxies
  in the $9'\times 9'$ box. For galaxies in the range $0.35 < z <
  0.55$ (dashed lines), we build the $r$-band cumulative LF of the
  group members (lower panel). The lens is the brightest galaxy in the
  group, most of the other members are fainter than 21\fm5.}
\end{center}
\end{figure*}

The left panel of Fig.~\ref{fig:image} shows a $g,r,i$ composite image.
Most of the faint objects in the field of view are galaxies, but the
environment is clearly not that of a rich cluster. The inset shows a
$16^{\prime\prime} \times 16^{\prime\prime}$ cut-out, in which the
central lens galaxy is surrounded by a $\sim 300^\circ$ ring of radius
$\sim 5^{\prime\prime}$. This makes it the largest, and one of the
most complete, optical rings ever discovered.

We obtained imaging follow-up data at the 2.5m {\it Isaac Newton
  Telescope} (INT), La Palma and spectroscopy
at the 6m BTA telescope of the {\it Special Astrophysical Observatory}
(SAO), Nizhnij Arkhyz, Russia. Observations were carried on the INT on
the night (UT) of 2007 May 12 with the Wide Field Camera (WFC).  The
exposure times were 600 s in each of the three wavebands $u, g$ and
$i$ -- which are similar to the SDSS filters.  The measured seeing
(FWHM) on the images ($0.33^{\prime\prime}$ pixels) was
$1.30^{\prime\prime}$, $1.26^{\prime\prime}$ and $1.21^{\prime\prime}$
in $u, g$ and $i$ respectively. The INT data are roughly a magnitude
deeper than the SDSS data and were reduced using the CASU INT WFC
pipeline toolkit~\citep{Ir01}.  The bottom right panel of
Fig.~\ref{fig:image} shows the $u,g,i$ composite field of view of
$24''\times 24''$ centered on the lens galaxy.  The \name\ is shown
with great clarity in the panels of Fig~\ref{fig:int}. We can extract
the properties of the LRG, such as magnitude, effective radius,
ellipticity and orientation, by masking out the ring emission and
fitting a PSF-convolved de Vaucouleurs profile as listed in
Table~\ref{tab:struct}. Our INT magnitudes agree with the SDSS
magnitudes reported in the Table, although SDSS overestimates the $g$
band effective radius because of contamination from the ring. The
shape of the isophotes of the LRG is shown in dotted lines. In the
right panel, the light from the lens galaxy is subtracted to leave a
clearer picture of the ring in the $g$ band. The surface brightness
profile along the ring in magnitudes arcsec$^{-2}$ is shown in the
inset. There are four maxima, A, B, C and D, whose right ascension and
declination offsets from the LRG are: A : ($3\farcs0,4\farcs6$), B :
($-1\farcs1, 5\farcs2$), C : ($-4\farcs7,2\farcs2$) and D :
($2\farcs0,-4\farcs0$) together with errors of $\lesssim 0\farcs4$.
There is some evidence that C may even be considered as two merging
images at C$_1$ ($-4\farcs7,2\farcs2$) and C$_2$
($-4\farcs8,-1\farcs7$).  Fig~\ref{fig:nbhd} shows the number density
of galaxies with photometric redshifts provided by SDSS in the range
$0.35 < z < 0.55$. In the left panel, a large-scale filamentary
structure can be discerned. The middle panel shows that the \name\
lies in a group of galaxies -- the enhancement in number density over
the background is $\sim 6$.  The lens is the brightest object in the
group of $\sim 26$ members, as is clear from the cumulative luminosity
function in the right panel.

Long-slit spectral observations were performed on 2007 May 15/16 with
the multi-mode focal reducer SCORPIO \citep{2005AstL...31..194A}
installed at the prime focus of the BTA~6-m telescope at the SAO.  The
seeing was $1\farcs7$.  A 1\farcs0 wide slit was placed to intercept
the two brighter arcs in the ring (C and D in Fig.~\ref{fig:int}) and
to include some of the light from the lens galaxy, as shown in the top
right panel of Fig.~\ref{fig:image}.  We used the VPHG550G grism which
covers the wavelength interval 3650--7550\,\AA\ with a spectral
resolution 8-10\,\AA\ FWHM.  With a CCD EEV 42-40 2k\,$\times$\,2k
detector, the reciprocal dispersion was $1.9$\,\AA\ per pixel.  The
total exposure time was 3600\,s, divided into six 10-minute
exposures. The target was moved along the slit between exposures to
ease background subtraction and CCD fringes removal in the data
processing.  The bias subtraction, geometrical corrections, flat
fielding, sky subtraction, and calibration to flux units
($F_{\lambda}$) was performed by means of IDL-based software.

\begin{figure*}
\begin{center}
\includegraphics[height=5.5cm]{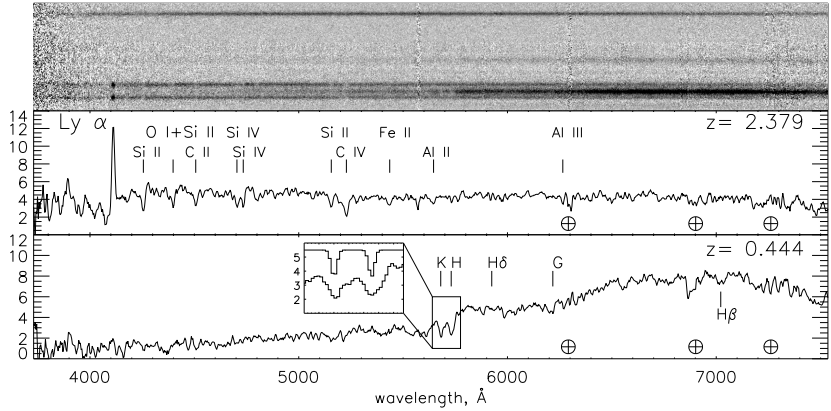}
\caption{\label{fig:spectrum} Top: Cutout of the SCORPIO 2D spectrum,
  the horizontal coordinate is the dispersion, the vertical coordinate
  is the location on the slit. In the lower part, 2 ring images are
  clearly visible at short wavelength (note the bright Ly$\alpha$
  blobs) with the lens appearing at longer wavelengths. Middle: Sum of
  two extracted 1D image spectra with absorption lines from Table 1 of
  Shapley et al. (2003). Bottom: 1D lens spectrum with Ca H and K
  lines marked. As a demonstration that the lines are resolved, we
  show in the inset a zoom of the H and K lines (lower) and the
  instrumental resolution (upper).  Note the prominent atmospheric
  absorption marked by $\oplus$ symbols.  The spectra are shown in
  flux units of $10^{-18}$ erg\,s$^{-1}$\,cm$^{-2}$\,$\AA^{-1}$.  }
\end{center}
\end{figure*}

The top panel of Fig.~\ref{fig:spectrum} shows a cut-out of the
two-dimensional spectrum with position along the slit plotted against
the dispersion.  The slit also passes through a nearby star, which
causes the spectrum in the topmost pixels. In the lower part, the blue
spectrum is dominated by two images of the source, whilst the red
spectrum by the lensing galaxy. The lower panels show extracted
one-dimensional spectra. The middle one is the sum of the two source
images; there is a strong narrow line which is Ly $\alpha$ emission,
together with accompanying Ly $\alpha$ forest bluewards and multiple
absorption lines redwards. This yields a measurement of the source
redshift as $z = 2.379$.  The lower panel is the lens galaxy spectrum,
which shows the characteristic features of a LRG.  The lens redshift
is $z = 0.444$. Although Ca H and K absorption is detected in the
lensing galaxy spectrum, the signal-to-noise (S/N) ratio is modest,
$\sim10$, and the resolution relatively low.  However, the inset in
the lower panel shows the instrumental resolution and the Ca H and K
lines, which are clearly resolved.  Performing fits of Gaussian line
profiles to the absorption produces a velocity dispersion estimate of
430$\pm$50\,kms$^{-1}$, where the principal uncertainty arises from
the placement of the `continuum'. The spectrograph slit was not
aligned across the centre of the galaxy but, given the relatively poor
seeing, the spectrum is dominated by light from within the half-light
radius of the galaxy. 

\begin{figure*}
\begin{center}
\includegraphics[height=5.cm]{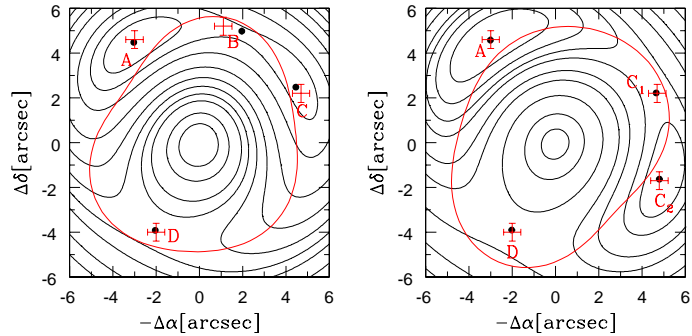}
\caption{\label{fig:model} Contours of the Fermat time delay surface
  for two possible lens models of \name, together with the locations
  of the stationary points which mark the predicted image positions.
  The critical curve of the lens model, which is also a contour of
  constant convergence, is shown in red, together with the observed
  image locations. Left: The model uses eqn. (5) of \citet{Ev03} with
  the Fourier coefficients ($a_0 = 9.89$, $a_2 = 0.090$, $b_2 =
  -0.11$, $a_3 = 0.02$, $b_3 = -0.04$) to reproduce image locations
  A,B, C and D.  Right: A similar model, but with Fourier coefficients
  ($a_0 = 10.07$, $a_2 = 0.066$, $b_2 = -0.22$, $a_3 = -0.03$, $b_3 =
  -0.01$) to reproduce image locations A, C$_1$, C$_2$ and D.}
\end{center}
\end{figure*}

\section{Discussion}

\subsection{Source}

The spectrum in Fig.~\ref{fig:spectrum} shows the source is a
star-forming galaxy at $z = 2.379$.  From the observed wavelengths of
the ten labelled absorption lines, we deduce a mean redshift $\langle
z_{\rm abs} \rangle = 2.3767 \pm 0.0006$, while the peak of the
Ly$\alpha$ emission line gives $z_{\rm em} \simeq 2.3824$. The overall
character of the spectrum is typical of BX galaxies in the surveys by
Steidel et al. (2004). These are galaxies at a mean redshift $\langle
z \rangle \simeq 2.2$ selected from their blue rest-frame UV colours.
In finer detail, the spectrum resembles most closely the subset of
these galaxies which are relatively young, with assembled stellar
masses $\langle M^{\ast} \rangle \simeq 5 \times 10^9 M_\odot$ and
metallicities of about 1/3 solar. The composite spectrum of galaxies
with these characteristics has been discussed by Erb et al. (2006a)
and has typical rest-frame equivalent widths of the interstellar lines
$W_{\rm IS} \simeq 1.5 - 2$\,\AA, and a similar strength of the Ly
$\alpha$ emission line.  The closest local analogue is the field
spectrum of nearby starburst galaxies (Chandar et al. 2005). The
difference between Ly$\alpha$ emission and interstellar absorption
redshifts found here is typical of high redshift star-forming galaxies
and is generally interpreted as resulting from large-scale outflows of
the interstellar medium in galaxies with high rates of star formation,
driven by kinetic energy deposited by massive star winds and
supernovae.  Adopting the median blueshift of 165\,km~s$^{-1}$ of the
interstellar absorption lines relative to the H\,{\sc ii} regions
producing H$\alpha$ emission (Steidel et al. 2007, in preparation), we
deduce a systemic redshift of $z_{\rm sys} = 2.379$.

The galaxy appears to be of fiducial luminosity. Interpolating between
the measured $g$ and $i$ magnitudes in Table 1, we deduce an absolute
magnitude at 1700\,\AA\ AB$_{\rm 1700} = -25.4$ in the standard
cosmology. If the magnification factor is $\sim 35$ (see next
Section), or 3.9 mag, this corresponds to an intrinsic AB$_{\rm 1700}
= -21.5$, or $L \simeq 1.6 L^{\ast}$, according to the recent
determination of the luminosity function of BX galaxies by Reddy et
al. (2007).  The colours of the lensed galaxy are typical of those of
most BX galaxies.  The $u\!-\!g$ and $g\!-\!i$ colours indicated by
the photometry in Table 1 imply a UV spectral slope redder than the
essentially flat spectrum ($F_{\nu} \propto \nu^0$) expected for an
unobscured star-forming galaxy (e.g. Leitherer et al. 1999).  Assuming
that the Calzetti et al. (2000) obscuration law applies, we deduce
$E(B\!-\!V) = 0.2$, close to the median of the distribution of the
values reported by Erb et al. (2006b) for BX galaxies. The
corresponding attenuation at 1700\,\AA\ is a factor of $\sim 6$.


\subsection{Lens}

\cite{Be06} found 70 galaxies with dispersions $> 350$ kms$^{-1}$ that
were not superpositions in the spectroscopic part of the SDSS
DR1. These are the galaxies with largest velocity dispersions and
might harbour the most massive black holes.  The fact that the
PSF-convolved de Vaucouleurs model gives an excellent fit to the light
distribution of the lens galaxy minimises the chance that the high
velocity dispersion is a product of superposition in our case.  The
lens is detected in the NVSS and FIRST surveys with an integrated flux
density at 20cm of 4.8 and 5.4mJy respectively.  Assuming a radio
spectrum of the form $S_\nu \propto \nu^\alpha$ ($\alpha=-0.7$) the
monochromatic radio power is $\rm 3.2 \times 10^{24}$ W Hz$^{-1}$
similar to the radio galaxies studied at $z \sim 0.7$ in the 2SLAQ
luminous red galaxy survey (Sadler et al.  2006). Of course, we have
assumed that all of the radio flux comes from the lens. In the nearby
Universe such powerful radio sources are associated with active
galactic nuclei rather than star-forming galaxies.

The $r$-band absolute magnitude of the lens is -23.45 at $z=0$.  This
assumes the SDSS $r$-band model magnitude of $r$=19.00, together with
the standard cosmology, a $k$ correction of $-0\fm87$, and the passive
evolution model of $+0\fm38$ \citep{Be03}.  This puts the lens in the
brightest bin for LRGs. The high luminosity is also indicated by the
red color ($g\!-\!i > 2.6$) of the galaxy. Color and luminosity also
correlate with velocity dispersion and mass (Figures 4 and 7 of
Bernardi et al. 2003). All these measurements support the idea that
the lensing galaxy is a very massive object.

Let us model the lens as a singular isothermal sphere galaxy with a
velocity dispersion $\sigma_{v}=430$ km\,s$^{-1}$.  For a lens
redshift of $0.44$ and a source redshift of $2.38$, the deflection due
to an isothermal sphere is $\sim 3.7^{\prime\prime}$.  As the LRG is
so massive, it provides most of the deflection needed. In physical
units, the ring radius is at a projected distance of $\sim 30$ kpc
from the center of the LRG.  The (cylindrical) mass enclosed within
the Einstein ring is $\sim 5.4 \times 10^{12}$ M$_\odot$. The
magnification can be estimated assuming that the source size is $\sim
0\farcs4$ ~\citep{La07}. The ratio of the area subtended by the ring
to that subtended by the source is $\sim 4R/ \delta r$, where $R$ is
the ring radius and $\delta r$ is the source size which is roughly
same as the ring thickness. This gives a magnification of $\sim 50$.
Though the lens galaxy provides most of the deflection, there is
probably a modest contribution from the environment. \cite{Ke01}
showed that the ellipticity of an Einstein ring is proportional to the
external shear. The \name\ is nearly a perfect circle.  Any
contribution from the galaxy group must therefore be modest. This is
surprising, as all other large separation lenses have a significant
contribution from the environment.

The ring has at least four density knots, whose locations are noted in
Section 2.  A more sophisticated algorithm that fits to the image
locations and relative brightnesses is provided by the method of
\citet{Ev03}. Here, the lens density has an isothermal profile in
radius, but the angular shape of the isodensity contours is given by a
Fourier series.  Fermat surfaces and critical curves are presented for
two possible models in Figure~\ref{fig:model}.  In the left panel, the
positive parity images are A and C, whilst the negative parity images
corresponding to saddle-points on the Fermat surface and are B and D.
In the right panel, C is regarded as a merging pair (C$_1$ and C$_2$),
whilst A and D are retained as images and B is discarded. In both
cases, the mass enclosed within the Einstein ring is $\sim 6 \times
10^{12}$ M$_\odot$, similar to our crude estimates, while the
magnification is in the range $25-35$.  Also possible is that the
\name\ is a sextuplet system, with C a conglomeration of three merging
images in addition to A,B and D (see e.g., Evans \& Witt 2001).

The combination of high absolute luminosity and large magnification
factor makes the \name\ the brightest galaxy known at $z > 2$. The
lens galaxy is one of the most massive LRGs ever detected. Detailed
studies of this remarkable system at a variety of wavelengths, from
optical to sub-mm will help us probe the physical nature of star
formation in the young universe, whilst detailed modeling will enable
us to study the interplay between baryons and dark matter in very
massive galaxies.

\acknowledgments The authors acknowledge with gratitude the support of
the EC 6th Framework Marie Curie RTN Programme MRTN-CT-2004-505183
("ANGLES"). The paper was partly based on observations collected with
the 6m telescope of the Special Astrophysical Observatory (SAO) of the
Russian Academy of Sciences (RAS) which is operated under the
financial support of Science Department of Russia (registration number
01-43). A.V.M. also acknowledges a grant from the President of Russian
Federation (MK1310.2007.2). Funding for the SDSS and SDSS-II has been
provided by the Alfred P.  Sloan Foundation, the Participating
Institutions, the National Science Foundation, the U.S. Department of
Energy, the National Aeronautics and Space Administration, the
Japanese Monbukagakusho, the Max Planck Society, and the Higher
Education Funding Council for England. The SDSS Web Site is
http://www.sdss.org/. The SDSS is managed by the Astrophysical
Research Consortium for the Participating Institutions. The
Participating Institutions are the American Museum of Natural History,
Astrophysical Institute Potsdam, University of Basel, Cambridge
University, Case Western Reserve University, University of Chicago,
Drexel University, Fermilab, the Institute for Advanced Study, the
Japan Participation Group, Johns Hopkins University, the Joint
Institute for Nuclear Astrophysics, the Kavli Institute for Particle
Astrophysics and Cosmology, the Korean Scientist Group, the Chinese
Academy of Sciences (LAMOST), Los Alamos National Laboratory, the
Max-Planck-Institute for Astronomy (MPIA), the Max-Planck-Institute
for Astrophysics (MPA), New Mexico State University, Ohio State
University, University of Pittsburgh, University of Portsmouth,
Princeton University, the United States Naval Observatory, and the
University of Washington.

\clearpage

\begin{deluxetable}{lll}
\tablecaption{Properties of the \name}
\tablewidth{0pt} \tablehead{ \colhead{Component} &
  \colhead{Parameter} & {~~~ } }
\startdata Lens & Right ascension & 11:48:33.15 \\
\null & Declination & 19:30:03.5 \\
\null & Redshift, $\zl$ & 0.444 \\
\null & Magnitudes (SDSS), $g_{\rm L}$, $r_{\rm L}$, $i_{\rm L}$ & 20\fm8,
19\fm0, 18\fm2 \\
\null & Effective radii (INT), $R_{\rm eff, g}$, $R_{\rm eff, i}$ &
$2.2^{\prime\prime}$, $1.7^{\prime\prime}$ \\
\null & Axis ratio (INT, in $g,i$) & 0.8, 0.9 \\
\null & Position angle (INT, in $g,i$) & $99^\circ$, $95^\circ$ \\
\null & Radio Flux (FIRST,NVSS) & $5.4$ mJy, $4.8$ mJy \\
Source & Redshift, $\zs$ & 2.379 \\
Ring & Diameter & $10\farcs2$ \\
\null & Length & $300^\circ$ \\
\null & Total magnitudes (INT) $u,g,i$ & 21\fm6, 20\fm1, 19\fm7 \\
\null & Mass Enclosed & $5.4 \times 10^{12}$ M$_\odot$ \enddata
\label{tab:struct}
\end{deluxetable}

\end{document}